\documentclass[11pt]{article}    
\usepackage[top=1in,bottom=1in,left=1in,right=1in]{geometry}    
\usepackage{diagbox}
\usepackage{multirow}
\usepackage[hidelinks]{hyperref}
\usepackage[all]{hypcap}
\usepackage{graphicx}
\usepackage{setspace}
\usepackage[parfill]{parskip}   
\usepackage{amsmath}
\usepackage{longtable}
\usepackage[usenames,dvipsnames]{color}
\usepackage{lscape}
\usepackage[toc,page]{appendix}
\usepackage{amssymb}
\usepackage{lineno}
\usepackage{fancyhdr}
\usepackage{lastpage}
\usepackage{arydshln}
\usepackage[font=small,labelfont=bf]{caption}
\usepackage{authblk}
\usepackage{soul}
\usepackage{lscape}
\usepackage{floatrow}
\usepackage{pdflscape}
\usepackage{amsfonts}
\usepackage{bbm}
\usepackage{array}
\usepackage{float}
\usepackage[export]{adjustbox}
\usepackage{subcaption}

\usepackage[numbers, square, comma, sort&compress]{natbib}

\newcommand{\rootpcp}{$\sqrt{PCP}$ }
\newcommand{\rootpcpc}{$\sqrt{PCP}$}

\newcommand{\ncpcp}{$nc\sqrt{PCP}$ }
\newcommand{\ncpcpc}{$nc\sqrt{PCP}$}

\usepackage[utf8]{inputenc}

\title{Principal Component Pursuit for Pattern Identification in Environmental Mixtures}

\author[1]{Elizabeth A. Gibson}

\author[2]{Junhui Zhang}
\author[2]{Jingkai Yan}
\author[1]{Lawrence Chillrud}
\author[1]{Jaime Benavides}
\author[1]{Yanelli Nunez}
\author[1]{Julie B. Herbstman}
\author[3]{Jeff Goldsmith}
\author[2]{John Wright}
\author[1,$\star$]{Marianthi-Anna Kioumourtzoglou}

\affil[1]{Department of Environmental Health Sciences, Columbia University Mailman School of Public Health, New York, NY}
\affil[2]{Department of Electrical Engineering, Columbia University Data Science Institute, New York, NY}
\affil[3]{Department of Biostatistics, Columbia University Mailman School of Public Health, New York, NY}
\affil[$\star$]{Corresponding author. 722 W. 168th Street, New York, NY, 10032, mk3961@cumc.columbia.edu}

\begin{document}

\maketitle
\clearpage

\section{Abstract}

\paragraph{Background and Aims:} Environmental health researchers often aim to identify sources or behaviors that give rise to potentially harmful environmental exposures. We have adapted principal component pursuit (PCP)---a robust and well-established technique for dimensionality reduction in computer vision and signal processing---to identify patterns in environmental mixtures. PCP decomposes the exposure mixture into a low-rank matrix containing consistent patterns of exposure across pollutants and a sparse matrix isolating unique or extreme exposure events.

\paragraph{Methods:} We adapted PCP to accommodate non-negative data, missing data, and values below a given limit of detection (LOD). We simulated data to represent environmental mixtures of two sizes with increasing proportions $<$LOD and three noise structures. We applied PCP-LOD to evaluate its performance compared with principal component analysis (PCA). 

We next applied PCP-LOD to an exposure mixture of 21 persistent organic pollutants (POPs) measured in 1,000 U.S. adults from the 2001--2002 National Health and Nutrition Examination Survey (NHANES). We applied singular value decomposition to the estimated low-rank matrix to characterize the patterns.

\paragraph{Results:} PCP-LOD recovered the true number of patterns through cross-validation for all simulations; based on an \textit{a priori} specified criterion, PCA recovered the true number of patterns in 32\% of simulations. PCP-LOD achieved lower relative predictive error than PCA for all simulated datasets with up to 50\% of the data $<$LOD. When 75\% of values were $<$LOD, PCP-LOD outperformed PCA only when noise was low.

In the POP mixture, PCP-LOD identified a rank-three underlying structure and separated 6\% of values as extreme events. One pattern represented comprehensive exposure to all POPs. The other patterns grouped chemicals based on known structure and toxicity.

\paragraph{Discussion:} PCP-LOD serves as a useful tool to express multi-dimensional exposures as consistent patterns that, if found to be related to adverse health, are amenable to targeted public health messaging.

\section{Introduction}

To assess exposure to multiple chemicals simultaneously, researchers must consider the high dimensionality of environmental exposures and the complex correlation structure across chemicals. Environmental epidemiologists may turn to dimension reduction or variable selection methods to weaken (or eliminate) correlations within the exposure matrix. When researchers are interested in identifying patterns within environmental exposure mixtures, they often employ dimension reduction techniques \cite{gibson2019complex}. Research questions concerning pattern identification commonly aim to represent underlying sources or behaviors that give rise to multi-pollutant exposures. Interpretable results may prove actionable if identified patterns reveal preventable or modifiable circumstances that lead to exposure. Their identification enables better informed policies and targeted interventions.

Research questions concerning pattern identification in environmental mixtures usually involve unsupervised statistical techniques whose solutions are obtained independently of any outcomes. Researchers apply common methods, such as principal component analysis (PCA) and factor analysis, to describe the variability in correlated chemicals in terms of underlying (i.e., latent) components. PCA is the most common dimensionality reduction tool used to identify patterns in environmental mixtures \cite{pang16, mak14_unc, gibsonnunez, robinson2018urban, manzano2015tnf}, but it has several limitations. First, various selection criteria exist to choose the number of components selected as patterns, such as the first $k$ principal components that explain a certain amount of variance, all components with singular values greater than one, or the components whose variances appear to the left of an `elbow' in a scree plot \cite{jolliffe1986principal}.  However, there is no guarantee that these criteria will agree \cite{jolliffe2016principal}. This leaves the burden on the researcher to determine the appropriate number of components, which is often based on implicit assumptions that are not always explicitly stated. Further, PCA has no guarantee of an interpretable solution \cite{lever2017points}. Its identified components are orthogonal by design, while patterns of environmental exposures are almost certainly not, and its solution, both chemical loadings and individual scores, may contain negative numbers, while actual chemical concentrations cannot \cite{friedman2001elements}. Finally, as a least squares method, PCA is susceptible to outliers, which may severely influence the solution \cite{wold1987principal}. Researchers also regularly employ dimension reduction methods beyond PCA, such as factor analysis or non-negative matrix factorization (NMF), for pattern recognition in environmental mixtures; these techniques work a bit differently than PCA, but they have similar drawbacks or introduce new ones (e.g., non-negativity may produce identifiability problems).

When analyzing high-dimensional datasets, a major challenge is how to recover low-dimensional patterns from noisy, incomplete, or erroneous measurements \cite{gull1978image}. In environmental health, observations below the analytic limit of detection (LOD) provide an example of incomplete data. Depending on the laboratory, these observations may be marked as $<$ LOD and not reported, or they may be reported as measured with less certainty than those $>$ LOD \cite{helsel2005more, helsel2005nondetects}. Identification of exposure patterns in datasets with large proportions of observations $<$ LOD proves challenging \cite{epalod}. 

Traditional methods to handle observations $<$ LOD include single and multiple imputation, the most common implementation being imputation with LOD/$\sqrt{2}$ \cite{barr2006survey}. This method was proposed in 1990 as providing more accurate estimation of the mean and standard deviation than imputation with LOD/2 and improved computational efficiency over a maximum likelihood method \cite{hornung1990estimation}. However, predictive accuracy is not often the goal in environmental epidemiology, and computational speed is no longer a barrier to new methods. Furthermore, substitution of values $<$ LOD with a fixed value (e.g., LOD/$\sqrt{2}$), especially when some information is available, will impact the distribution of the data, potentially severely impacting exposure pattern identification in the study population \cite{helsel1990less}.

Here, we introduce a novel technique to identify patterns in environmental mixtures, adapting a robust and well-established method for data dimensionality reduction and pattern recognition in computer vision applications, principal component pursuit (PCP). PCP decomposes the exposure data matrix into a low-rank matrix (to identify underlying patterns of exposure across the pollutants) and a sparse matrix (to identify unusual, unique, or extreme exposure events) \cite{candes2011robust}. PCP has several advantages over PCA in the area of pattern detection in environmental mixtures. In a recent PCP extension, square root PCP (\rootpcpc), \citet{zhang2021square} derived a new formulation with a universal choice of regularization parameter. Thus, the user is not required to choose or tune hyperparameters. We combined this with a separate extension introducing a non-convex penalty on the low-rank matrix that performs well with data that may not have a strong underlying structure \cite{netrapalli2014non, chen2020bridging}. Estimation of the sparse matrix is especially advantageous. Traditional methods are sensitive to unusual or extreme exposure events; the patterns identified by PCP are not influenced by outlying values. Instead, exposures that are not explained by patterns in the low-rank matrix are separated in the sparse matrix and available for to the researcher.

To our knowledge, this is the first time that PCP has been considered in pattern identification in environmental health or epidemiology. Additionally, we have included three novel extensions designed uniquely for chemical mixtures: (1) a distinct penalty for observations $<$ LOD (PCP-LOD) that has improved distributional assumptions over single imputation and adapts to study-specific confidence in measurement, (2) a non-negativity constraint on the low-rank matrix to improve interpretability of results, and (3) procedures to accommodate missing values. We also implemented a cross-validation approach (see ~\ref{methods:stat}) so that the choice of estimated components is not as subjective as in other methods. In this work, we conducted a simulation study based on real multi-pollutant exposures, simulating an increasing proportion of observations measured $<$ LOD and varying levels and structure of added noise. We use these to compare PCP-LOD performance to that of PCA with values $<$ LOD imputed as LOD/$\sqrt{2}$. Finally, we applied PCP-LOD to an environmental health dataset of persistent organic pollutants (POPs) measured in the 2001--2002 cycle of the National Health and Nutrition Examination Study (NHANES) to identify consistent patterns of POP exposure while isolating unique or extreme events.

\section{Methods}

\subsection{Principal component pursuit}
\label{methods:stat}
We present PCP as a robust method for dimensionality reduction and pattern identification \cite{candes2011robust}. Given an exposure data matrix $X_{n,p}$, where $n$ is the number of observations and $p$ is the number of pollutants, PCP seeks to express $X$ as a superposition of two matrices: a low-rank matrix $L_{n,p}$ where $r = \mathrm{rank}(L) \ll \min(n,p)$, and a sparse matrix $S_{n,p}$ where most entries are zero. Because $L$ is of rank $r \ll p$, its rank can correspond to underlying patterns in exposure, such as specific sources or certain behaviors. $L$ is still defined in terms of the original variables, i.e., patterns are not directly estimated. PCP may be paired with various matrix factorization techniques (e.g., SVD, PCA, factor analysis, or NMF) to extract chemical loadings and individual scores. The rank of $L$ or the number or location of nonzero entries in $S$ do not need to be \textit{a priori} defined.

We incorporated two PCP extensions that suit features of environmental mixtures data. First, \citet{zhang2021square} recently proposed \rootpcp with a noise-independent universal choice of regularization parameters. Previous formulations of PCP required knowledge of the true noise level to determine the appropriate parameters \cite{zhou2010stable, chen2015fast, chen2020bridging}. This is problematic in environmental mixtures where we cannot know or accurately estimate the underlying noise level, and it would leave the researcher with the subjective task of tuning parameters on a per-dataset basis. Zhang et al. provide a more practical approach to pattern recognition in environmental mixtures.

As first proposed, PCP minimizes a weighted combination of the nuclear norm of $L$, $\|L\|_{*}$, and the $\ell^1$ norm of $S$, $\|S\|_{\ell^1}$ \cite{chandrasekaran2011rank,candes2011robust,zhou2010stable}. Notably, this formulation has the desirable quality of convexity, meaning that every local optimum is a global optimum. This, together with the particular structure of the $\ell^1$ and nuclear norms guarantees that the resulting optimization problem can be solved efficiently \cite{lin2010augmented,boyd2004convex,Wright-Ma-2021}. In practice, however, the nuclear norm assumes a stronger low-rank structure (i.e., slowly decaying singular values) than what is the case in many real-life environmental mixtures (e.g., POPs or air pollution). To address unsatisfactory performance with the nuclear norm, we replaced it with a rank-$r$ projection. While the nuclear norm is convex, the rank-$r$ projection is not. However, closely related non-convex formulations are accompanied by theoretical guarantees of equivalent performance with the convex implementation \cite{netrapalli2014non, chen2020bridging}. Combining a non-convex rank projection with \rootpcpc, we solve the following optimization problem:
\begin{equation}\label{eqn:ncvxroot}
\text{non-convex }\sqrt{PCP} := \hspace{0.5ex} \min _{L, S} \hspace{1ex} \mathbbm{1}_{\mathrm{rank}(L) \le r} + \lambda\|S\|_{1} + \mu\|L+S-X\|_{F},
\end{equation}
where $X$ denotes the original data matrix. The two parameters, $\lambda$ and $\mu$, are not tuned by the researcher; instead, they are each set using single universal values, $\lambda = 1/\sqrt{n}$ from \citet{candes2011robust}, and $\mu = \sqrt{p/2}$ from \citet{zhang2021square} which have been shown theoretically to yield near-optimal estimation performance. The indicator function $\mathbbm{1}_{\mathrm{rank}(L) \le r}$ constrains $L$ to be of rank $\le r$; the $\ell^1$ norm $\|S\|_{1}$ is the sum of the absolute values of the entries of $S$ and encourages $S$ to be sparse; the final term is the error between the predicted and the observed values, which favors a solution that is close to the original data.

\subsubsection{Environmental Health-Relevant Extensions} 

To better adapt non-convex \rootpcp (\ncpcpc) for use with environmental data, we extended this method in three ways. First, we modified the algorithm to allow for missing values. This proves beneficial to environmental datasets which often include participants with missing exposure measurements. It also enables the cross-validation procedure outlined in Section~\ref{methods:eval}. Next, we constrained the low-rank matrix to be non-negative. Non-negativity in $L$ allows for individual pattern scores and chemical loadings on patterns on the same support as the original chemical distributions. We tailored the third extension to observations $<$ LOD. We introduced a diverging penalty, $\psi_{\mathrm{LOD}}$, in the \ncpcp solution to accommodate values $<$ LOD when they are not available to the users, as is most commonly the case. This penalty treats all estimated values from zero to the LOD as equally good approximations (Equation~\ref{eqn:lod}, line 2), removing the error term from the objective function: 
\begin{equation}
    \text{PCP-LOD} := \min_{L \ge 0, \; S} \mathbbm{1}_{rank(L) \le r} + \lambda\|S\|_{1} + \mu \; \psi_{\mathrm{LOD}}(L+S-X) 
\end{equation}
with
\begin{equation}\label{eqn:lod}
\psi_{\mathrm{LOD}}(L+S-X) = \left( \sum_{ij} \begin{cases} (L_{ij}+S_{ij}-X_{ij})^2 
  & \quad \text{if } X_{ij} \ge \delta_{ij}, \\
  0
  & \quad \text{if } X_{ij} < \delta_{ij} \ \& \ 0 \le (L_{ij}+S_{ij}) \le \delta_{ij}, \\
  (L+S-\delta_{ij})^2 
  & \quad \text{if } X_{ij} < \delta_{ij} \ \& \ (L_{ij}+S_{ij}) > \delta_{ij}, \\
  (L+S)^2 
  & \quad \text{if } X_{ij} < \delta_{ij} \ \& \ (L_{ij}+S_{ij}) < 0, \\
  \end{cases} \right)^{1/2}.
\end{equation}

Here, $\delta$ is a matrix of LOD values, and $\delta_{ij}$ represents the observation-specific LOD. This is an attribute of the data specified by the researcher; it can be common across all chemicals, chemical-specific, or chemical- and individual-specific, depending on the measurements. If all observations are $>$ LOD, this equation simplifies to Equation~\ref{eqn:ncvxroot}. For estimated values $>$ LOD (Equation~\ref{eqn:lod}, line 3) or $<0$ (Equation~\ref{eqn:lod}, line 4), we include more stringent penalties than in Equation~\ref{eqn:ncvxroot}, which act to push estimates to the known range, while for observations $X_{ij}$ below the LOD (Equation~\ref{eqn:lod}, line 2), we place no penalty on estimated values $L_{ij} + S_{ij}$ lying between $0$ and the LOD. 

\subsection{Simulations} 
\begin{figure}
\centering
\begin{subfigure}[c]{0.55\textwidth}
\includegraphics[width=1\linewidth]{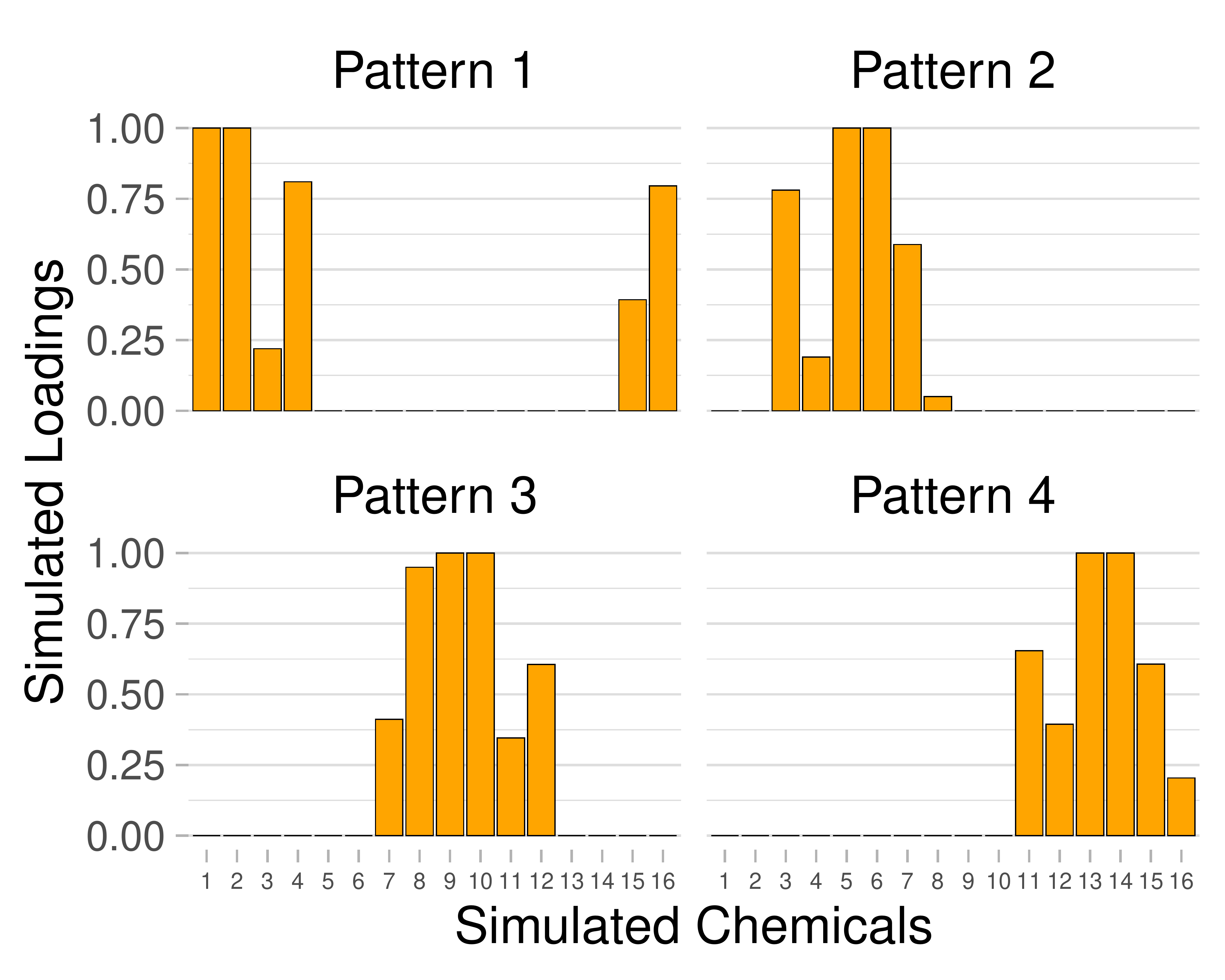}
\caption{Patterns}
\label{fig:simpatterns}
\end{subfigure}
\hspace{-1em}
\begin{subfigure}[c]{0.45\textwidth}
\includegraphics[width=.95\linewidth]{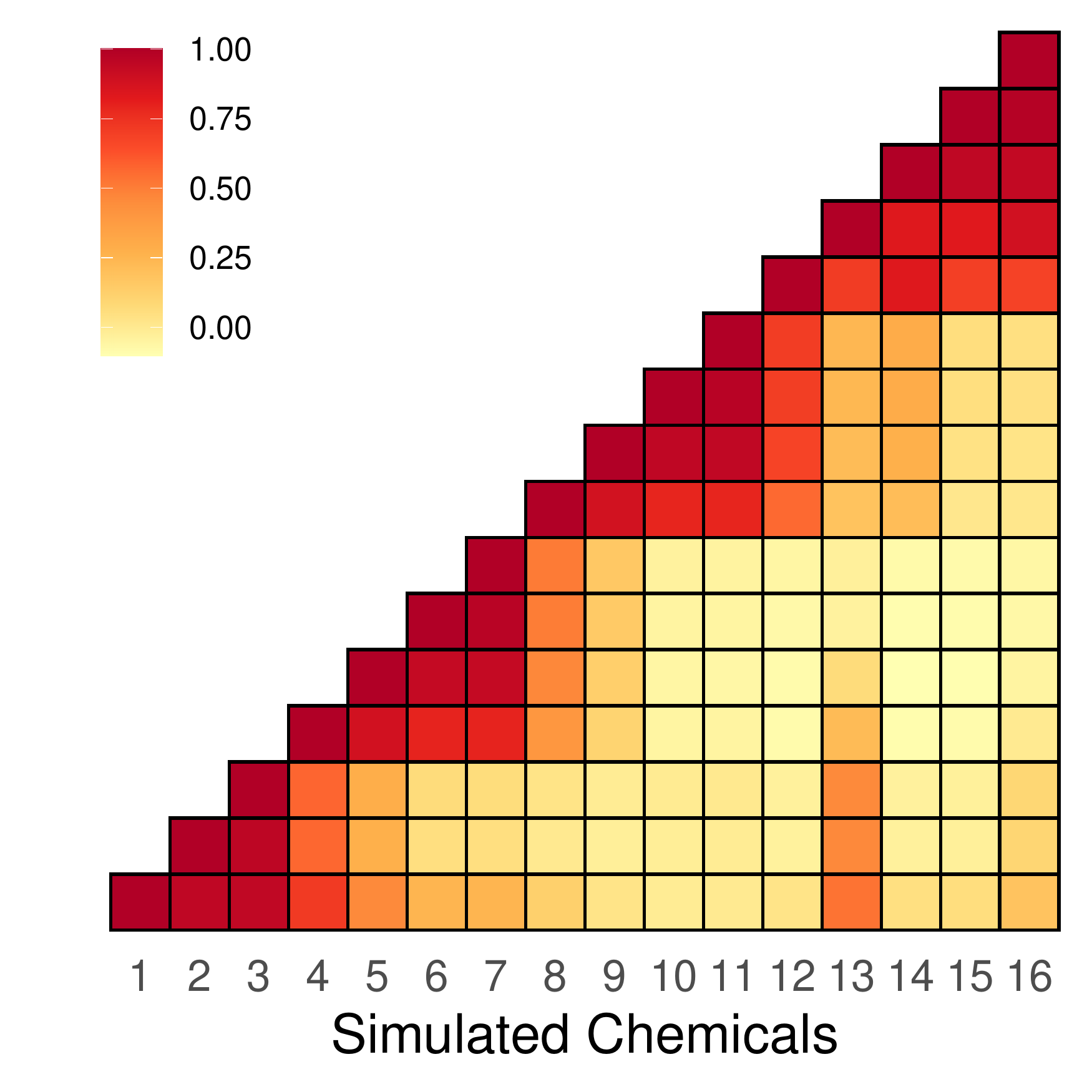}
\caption{Correlations}
\label{fig:simcorr}
\end{subfigure}
\hfill
\caption{\textbf{(a)} Representative simulated chemical loadings. We simulated 100 examples for mixture size $p$ = 16 (depicted here) and $p$ = 48. Here, two chemicals load solely on each of the four patterns. The remaining chemicals appear in two patterns each. \textbf{(b)} Correlation matrix of one simulated dataset ($p$ = 16) with high noise.}
\end{figure}

We simulated 100 exposure matrices for all combinations of two mixture sizes, three noise structures, and three detection proportions (1800 total). We generated datasets of 500 observations each, $\left\{\mathbf{x}_{i}\right\}_{i=1}^{500},$ where $\mathbf{x}_{i} = \left(x_{i,1}, \ldots, x_{i,p}\right)^{\mathrm{T}}$ presents an exposure profile with $p$ mixture components. We specified $r = 4$ underlying patterns and investigated two mixture sizes ($p = 16$ and $p = 48$). We first simulated chemical loadings ($r \times p$) to represent realistic environmental patterns where some chemicals were distinct to a single pattern and some chemicals appeared in multiple patterns. Each pattern included $\dfrac{p}{8}$ chemicals that loaded distinctly and $\dfrac{p}{4}$ chemicals that overlapped with a second pattern. Distinct chemicals were given a loading of 1 on the single pattern on which they loaded and a loading of 0 for the remaining patterns. One third of the chemicals appeared in only one pattern; two thirds of the chemicals appeared in two patterns. This design corresponds to multiple environmental sources giving rise to the chemicals in the mixture. Overlapping chemicals were drawn from a Dirichlet distribution so that their loadings would sum to 1 over all patterns. Of the four loadings across the four patterns for each chemical, two were drawn from $\operatorname{Dir}(\alpha_1=1, \alpha_2=1)$ and two were set to zero. This introduced variability into the overlapping chemical loadings (Figure~\ref{fig:simpatterns}).

We next generated individual scores ($n \times r$). We drew scores independently from $\log\mathcal{N}(\mu = 1, \sigma = 1)$. We created the simulated data from matrix products of individual scores by chemical loadings with added noise, replacing negative values with zero. We generated noise in one of three ways, (1) low Gaussian noise ($\mathcal{N}(0,1)$), (2) high Gaussian noise ($\mathcal{N}(0,5)$), (3) or low Gaussian noise with high sparse events. Figure~\ref{fig:simcorr} shows an example simulated correlation matrix. Finally, we designated a quantile (25$^{th}$, 50$^{th}$, or 75$^{th}$) and set all values below the threshold as $<$LOD.

\subsection{Study population} For pattern recognition in an environmental mixture with varying detection limits across chemicals, we chose a mixture of dioxins, furans, and polychlorinated biphenyls (PCBs) measured in U.S. adults from the 2001--2002 NHANES cycle. NHANES inclusion criteria have been reported previously \cite{zipf2013health}. For the chosen cycle, 11,039 participants were interviewed. One third of participants aged 12 years and older were eligible for environmental chemical analysis. We removed individuals below 18 years of age or without any POP measurements, resulting in a final study sample of 1,000. Eighteen PCBs, seven dioxins, and nine furans were measured. Exposure assessment of POPs in NHANES has been described previously \cite{cdc2002a, cdc2002b}. Of the POPs measured, 21 detected in at least 50\% of all samples were included in our analyses. All POP values were lipid-adjusted by the U.S. Centers for Disease Control and Prevention (CDC) \cite{akins1989estimation}. 

\subsection{Implementation \& Evaluation}
\label{methods:eval}
We determined the appropriate rank for PCP-LOD and the number of components to retain from PCA in the same manner for all experiments. For PCA, we \textit{a priori} defined our component retention criterion as the first $k$ components that explained $\geq 80\%$ of the variance in the data, as seen previously in environmental mixtures applications (e.g., \citet{gibsonnunez}). While it is possible to perform cross-validation on PCA \cite{diana2002cross, krzanowski1987cross}, it is not a common practice in applied environmental health research. For PCP-LOD we used the default parameters for $\lambda$ and $\mu$ and cross-validated to select the rank of the $L$ matrix. We set an initial grid of rank values from 1 to 10 for all scenarios. We performed this cross-validation approach on a single representative dataset for each combination of simulated mixture sizes ($p$ = 16 and 48), proportions $<$ LOD (25\%, 50\%, and 75\%), and noise structures (low, high, and sparse) and for the POP mixture.

To cross-validate PCP-LOD on a single dataset, $X$, we repeated the following steps 100 times for each rank $r \in [1,10]$. (1) We randomly corrupted 20\% of the mixture $X$ as missing (i.e., set the value to \texttt{NA}) to serve as a held-out test set, denoted $X_\Omega$, yielding the corrupted matrix $\Tilde{X}$. (2) We ran PCP-LOD on $\Tilde{X}$ to obtain $\hat{L}$ and $\hat{S}$. (3) We recorded the relative recovery error of $\hat{L}_\Omega + \hat{S}_\Omega$ compared with the observed data $X_\Omega$ in the held-out set, calculated via the Frobenius norm, $||X_\Omega - \hat{L}_\Omega - \hat{S}_\Omega ||_F\,/\,||X_\Omega||_F$. Finally, for each rank, we aggregated the average relative recovery error across 100 runs and chose the optimal rank, $\hat{r}$, as that with the lowest mean relative recovery error on the held-out set. We subsequently ran PCP-LOD on the full dataset $X$ with the selected rank $\hat{r}$.

We ran PCP-LOD and PCA on all simulated datasets. We compared PCP-LOD and PCA to assess their relative performance when faced with large proportions of non-detectable observations. For PCP-LOD we estimated the rank of $\hat{L}$, the sparsity of $\hat{S}$, and their relative change to assess stability of the solution across increasing proportions of data $<$ LOD. Because the sparse matrix may contain non-zero values so close to zero as to be considered zero, we set a threshold above which to regard values as legitimate extreme exposures. We evaluated sparse events two standard deviations of the model residuals ($\hat{S} + \hat{\varepsilon}$), per chemical, from zero, i.e., $2 \times \sqrt{\mathrm{Var}(X_{p}^{\mathrm{obs}} - \hat{L}_{p}^{\mathrm{obs}})}$, where $\mathrm{obs}$ indicates values above the LOD in the simulated data.

For both PCP-LOD and PCA, we calculated relative predictive error as the ratio of the error to the truth in terms of their Frobenius norm: $\left\lVert \mathrm{Truth} - \mathrm{Predicted}\right\rVert_F / \left\lVert  \mathrm{Truth}\right\rVert_F$. For PCP-LOD we interpreted $\hat{L}$ as the predicted values, and for PCA we constructed predicted values as the product of the score matrix (i.e., the coordinates of the rotated data on the principal components) by the rotation matrix (i.e., right eigenvectors), truncated at the chosen rank. We defined the `truth' as simulated values before noise or sparse events were added. Finally, we assessed the stability of the identified patterns using the relative prediction error of the singular value decomposition (SVD).

\subsection{Application}
Prior to the application to the NHANES POP mixture, we examined distributional plots and descriptive statistics for all variables. We scaled all exposure concentrations by their standard deviations to make variances comparable across chemicals. The solution, thus, cannot be influenced by high-variance pollutants. We used PCP-LOD to separate unique events from underlying patterns. Following PCP, we extracted individual scores and pattern loadings from $\hat{L}$ using SVD. We compared scores, loadings, and overall relative error with those obtained from PCA. We present unique events and interpret observed patterns. All analyses were conducted using R version 4.0.4 \cite{rrr}.

\section{Results}

\subsection{Simulations} 
\begin{figure}
    \centering
\includegraphics[width=.75\textwidth]{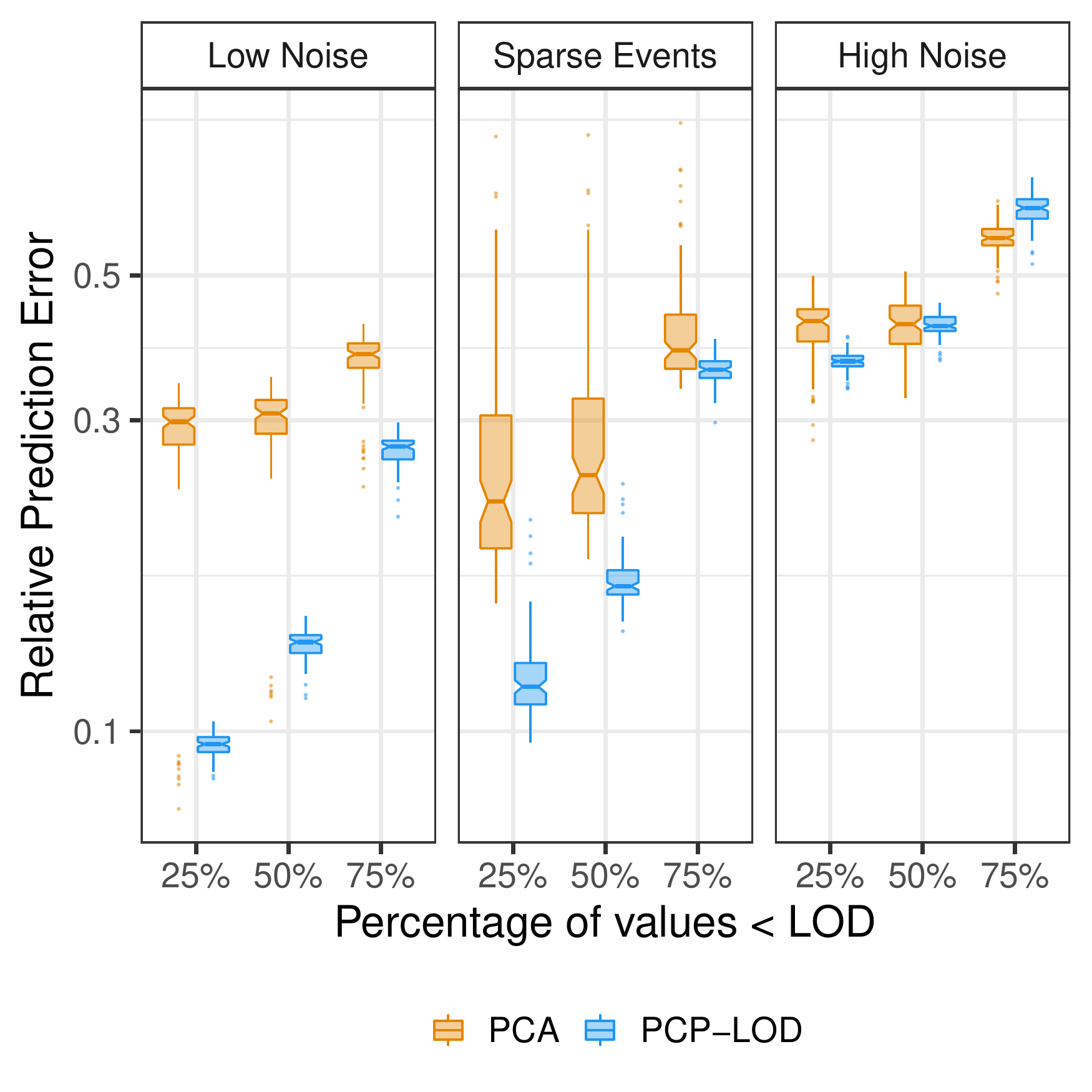}
   \caption{Overall relative predictive error of PCP-LOD and PCA on simulated data with $p$ = 16 across increasing proportions of data below the limit of detection. The panels show results for different structures of added noise. Box plots display summary statistics for each method across 100 simulations. The bottom and top hinges of the boxes correspond to the first and third quartiles (the 25$^{th}$ and 75$^{th}$ percentiles), respectively. The upper (lower) whiskers extend from the hinge to the largest (smallest) value no further than 1.5 $\times$ IQR from the hinge (where IQR is the inter-quartile range, or distance between the first and third quartiles).}
    \label{fig:overall}
\end{figure}

\begin{figure}
    \centering
\includegraphics[width=.75\textwidth]{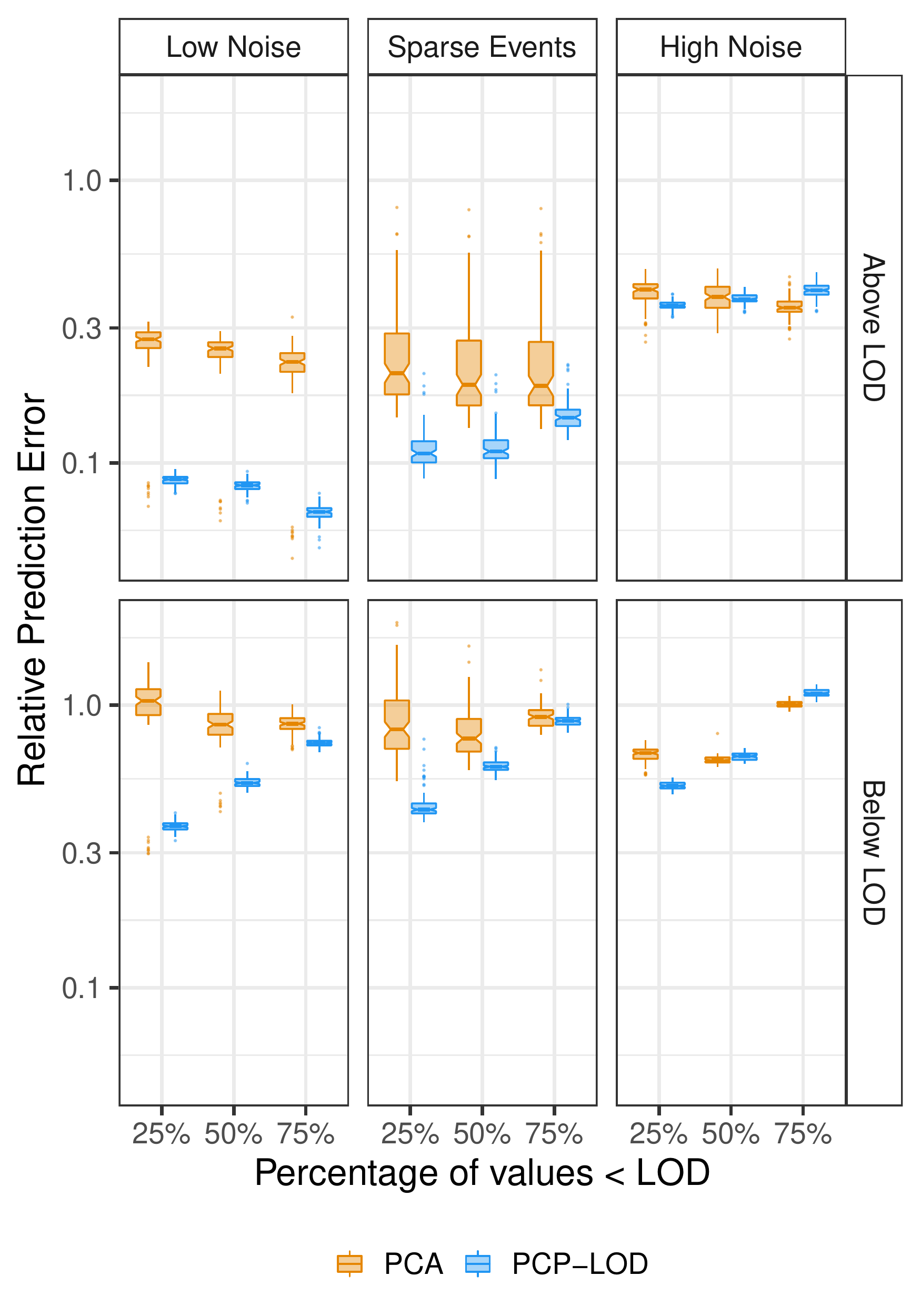}
   \caption{Relative predictive error of PCP-LOD and PCA on simulated data with $p$ = 16 stratified by detection. The panel columns separate results from different structures of added noise, and the panel rows separate values that were simulated as observed (top row) from those simulated as below the limit of detection (bottom row). Box plots display summary statistics for each method across 100 simulations.}
    \label{fig:above_below}
\end{figure}

We ran PCP-LOD and PCA on all simulated datasets. PCP-LOD had lower relative prediction error across the majority of  mixture sizes ($p$ = 16 and 48), proportions $<$ LOD (25\%, 50\%, and 75\%), and noise structures (low, high, and sparse). PCP-LOD outperformed PCA on all simulations with low noise, simulations with high noise with up to 50\% $<$ LOD, and simulations with low noise and added sparse events with up to 50\% $<$ LOD (Figure~\ref{fig:overall}). Figures~\ref{fig:overall}~and~\ref{fig:above_below} present simulations where $p$ = 16; corresponding figures where $p$ = 48 are included in Supplemental Figures~\ref{fig:overall_48} and \ref{fig:above_below_48}.

\begin{figure}
    \centering
\includegraphics[width=.75\textwidth]{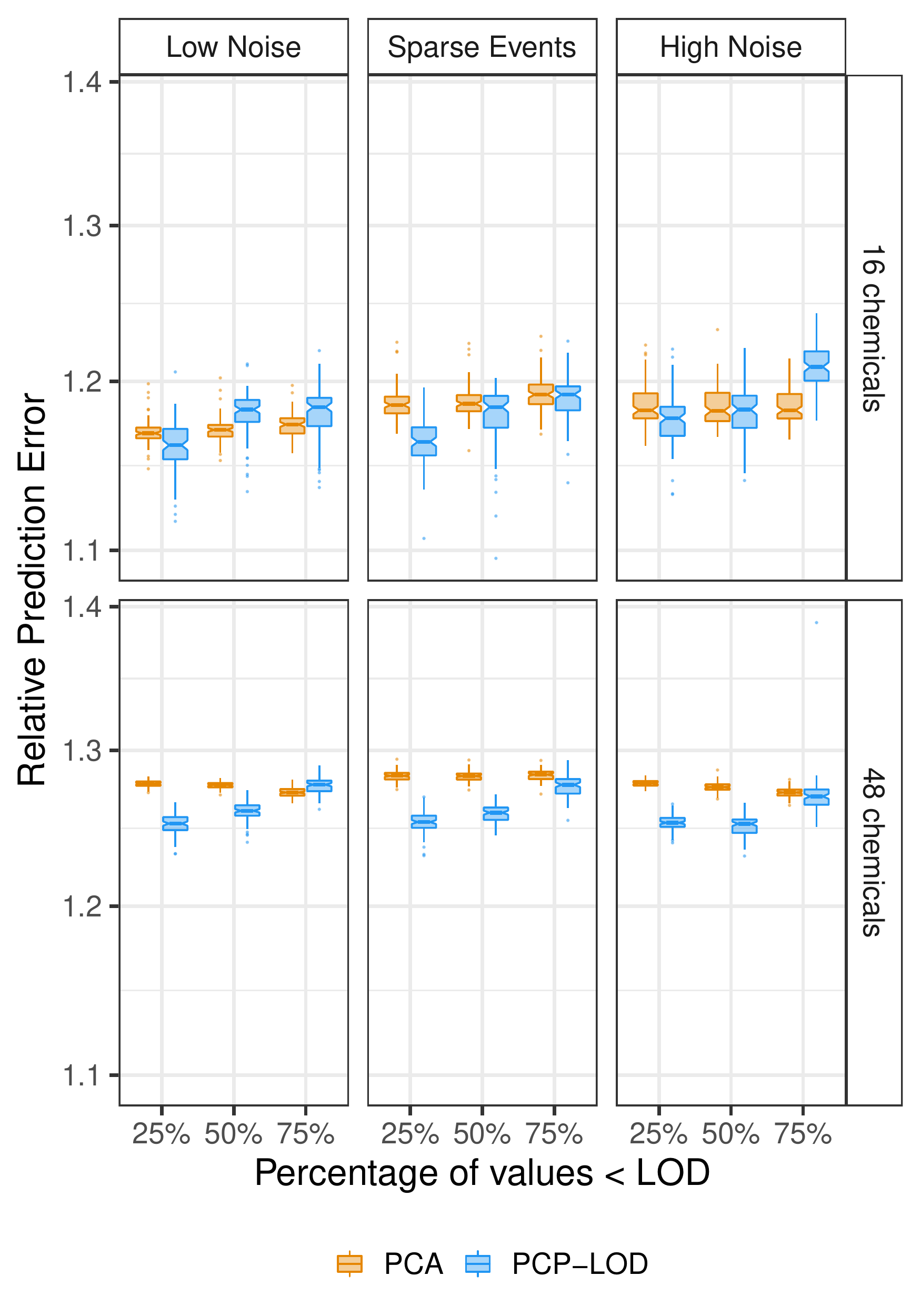}
   \caption{Relative predictive error of PCP-LOD and PCA solution scores (i.e., left eigenvectors) compared with those of the simulated data before noise was added. The panel columns separate results from different structures of added noise, and panel rows present two simulated mixture sizes. Box plots display summary statistics for each method across 100 simulations.}
    \label{fig:svd}
\end{figure}

PCP-LOD was more affected by the proportion of data $<$ LOD, which can be seen in the larger step size between box plots in Figure~\ref{fig:overall}. The decline in PCP-LOD predictive accuracy as the proportion of values $<$ LOD increased appears because of poorer performance on values $<$ LOD in high noise scenarios (Figure~\ref{fig:above_below}). Relative prediction error for values $>$ LOD was approximately constant for PCP-LOD and PCA. Supplemental Tables~\ref{table:supp1},~\ref{table:supp2}, and~\ref{table:supp3} contain the median and inter-quartile range (IQR) of relative error for predicted values overall and stratified by LOD.

Next, we assessed the stability of the identified patterns using the SVD of the simulated data before noise or sparse events were added and compared this with the SVD of the $\hat{L}$ matrix and of PCA results. Figure~\ref{fig:svd} depicts the relative prediction error comparing the left eigenvectors (comparable to scaled individual scores) of the PCP-LOD and PCA solutions with those of the simulated `truth.' PCP-LOD's median relative prediction error is generally lower than PCA's for the larger mixture size and higher than PCA's for the smaller mixture size. However, these patterns appear quite stable over increasing proportions of data $<$ LOD for both methods. PCP-LOD solutions achieved lower relative prediction error on chemical loadings (i.e., right eigenvectors) across all simulations (Supplemental Figure~\ref{fig:svd_right}).

Across PCP-LOD solutions, between 2\% and 10\% of $\hat{S}$ entries were non-sparse. We found decreasing sparsity as the proportion $<$ LOD increased, with 3\% (IQR: 2\%, 4\%), 6\% (IQR: 4\%, 7\%), and 7\% (IQR: 3\%, 8\%) unique events, on average, found in simulations with 25\%, 50\%, and 75\% $<$ LOD, respectively. For simulations that included sparse events in the noise structure, PCP-LOD correctly included 69\% (IQR: 67\%, 71\%), 70\% (IQR: 68\%, 72\%), and 65\% (IQR: 62\%, 67\%) of sparse values in the $\hat{S}$ matrix, on average for simulations with 25\%, 50\%, and 75\% $<$ LOD, respectively.

\subsection{Application} 
\begin{figure}
    \centering
\includegraphics[width=1\textwidth]{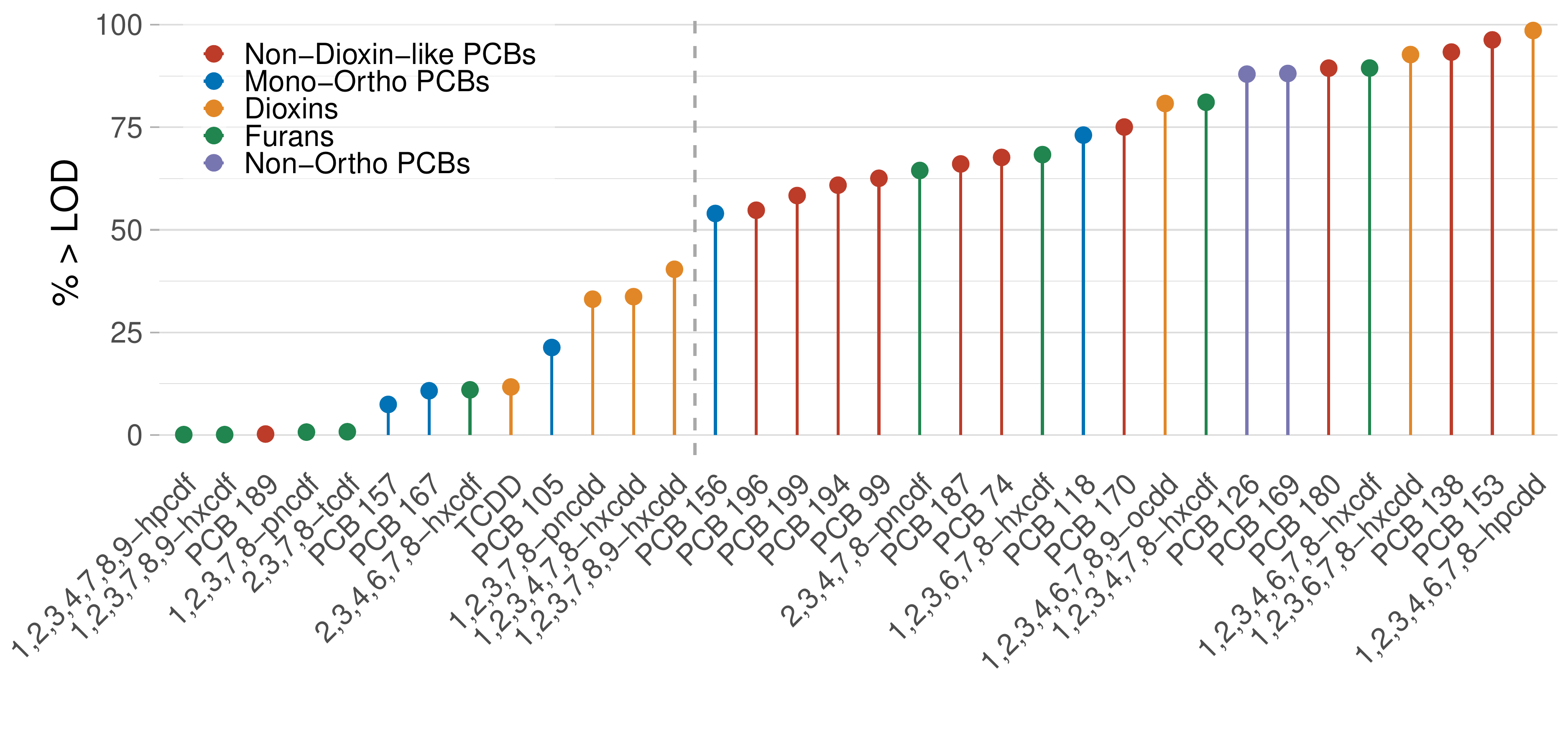}
   \caption{Detection frequency of persistent organic pollutants measured in NHANES 2001--2002. All congeners to the right of the vertical dashed line were detected in $>$ 50\% of samples and included in the analysis.}
    \label{fig:detect}
\end{figure}

\begin{figure}
\centering
\begin{subfigure}[b]{0.5\textwidth}
\includegraphics[width=1\linewidth]{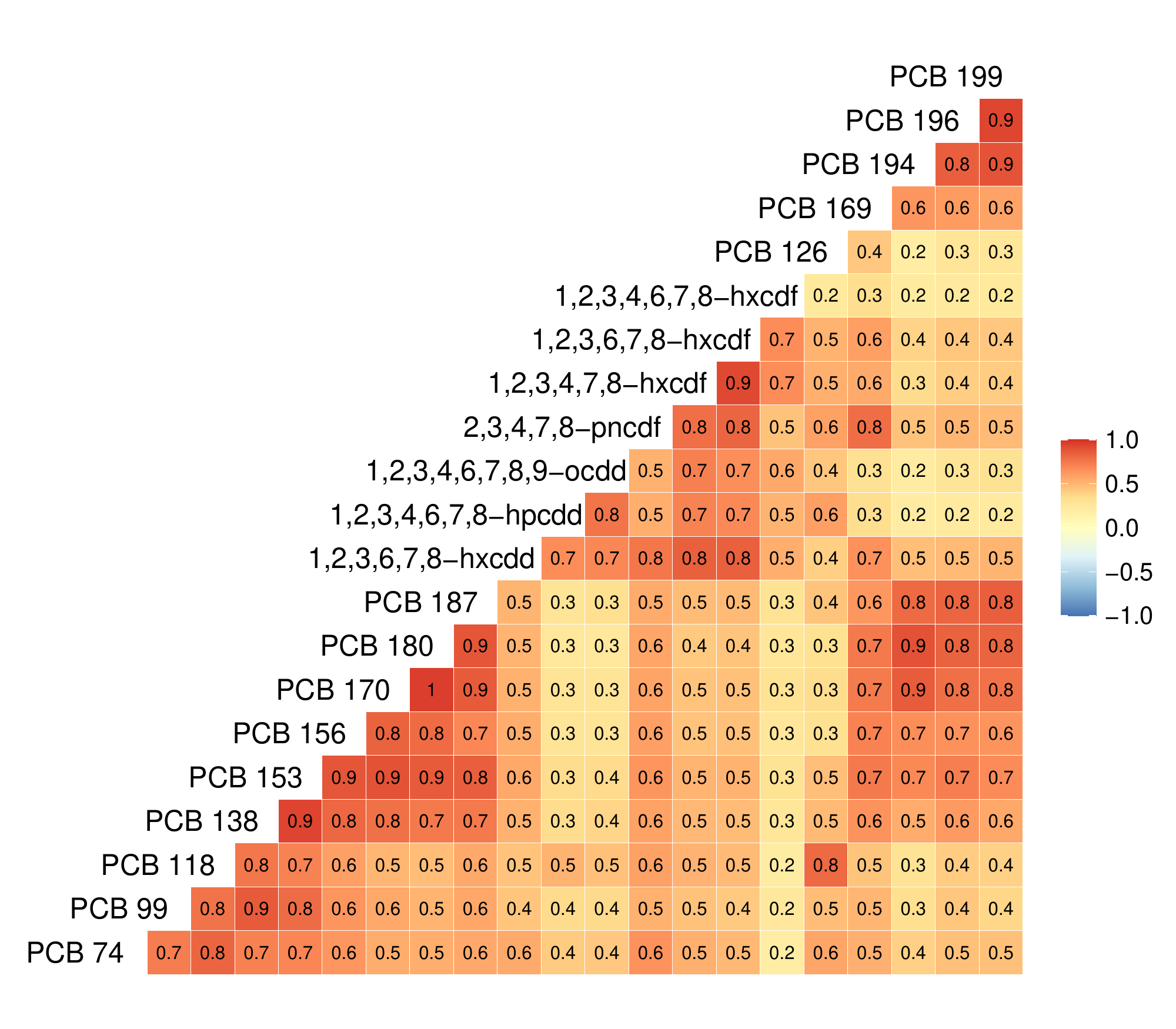}
\caption{Raw POP data}
\label{fig:nhanes_corr}
\end{subfigure}
\hspace{-1em}
\begin{subfigure}[b]{0.5\textwidth}
\includegraphics[width=1\linewidth]{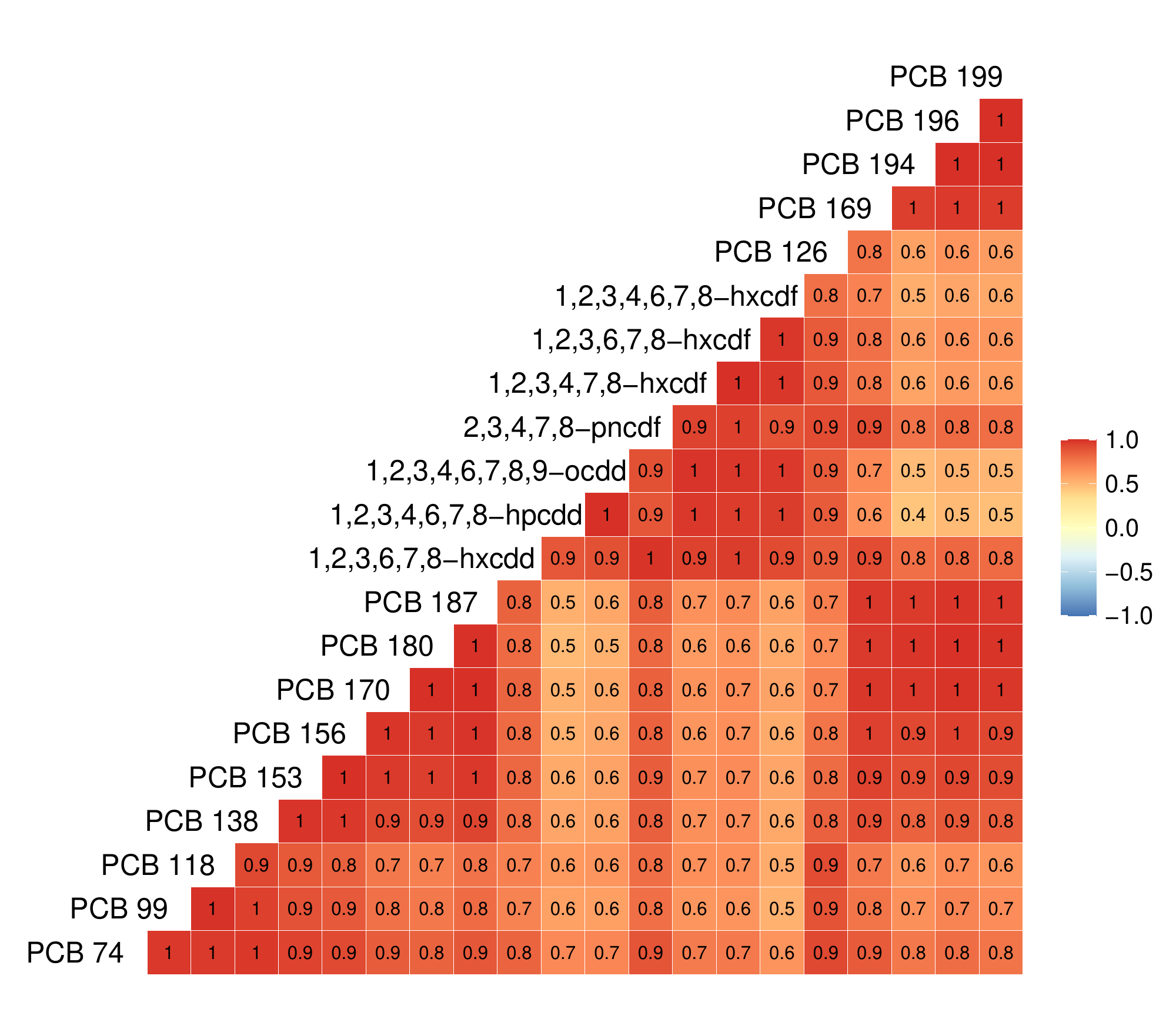}
\caption{$\hat{L}$ matrix}
\label{fig:nhanes_corr_l}
\end{subfigure}
\caption{\textbf{(a)} Spearman correlation matrix of 21 persistent organic pollutants measured in NHANES 2001--2002. Observations below the limit of detection were handled by case-wise deletion. \textbf{(b)} Spearman correlation matrix of low-rank structure across POPs estimated by PCP-LOD.}
\end{figure}

Thirty-four POPs were measured in the NHANES 2001--2002 cycle. Detection frequency is presented in Figure~\ref{fig:detect}. Fourteen PCBs, four furans, and three dioxins were detected in $>$ 50\% of samples. Exposure levels of POPs were all positively correlated (Figure~\ref{fig:nhanes_corr}).

\begin{figure}
    \centering
\includegraphics[width=1\textwidth]{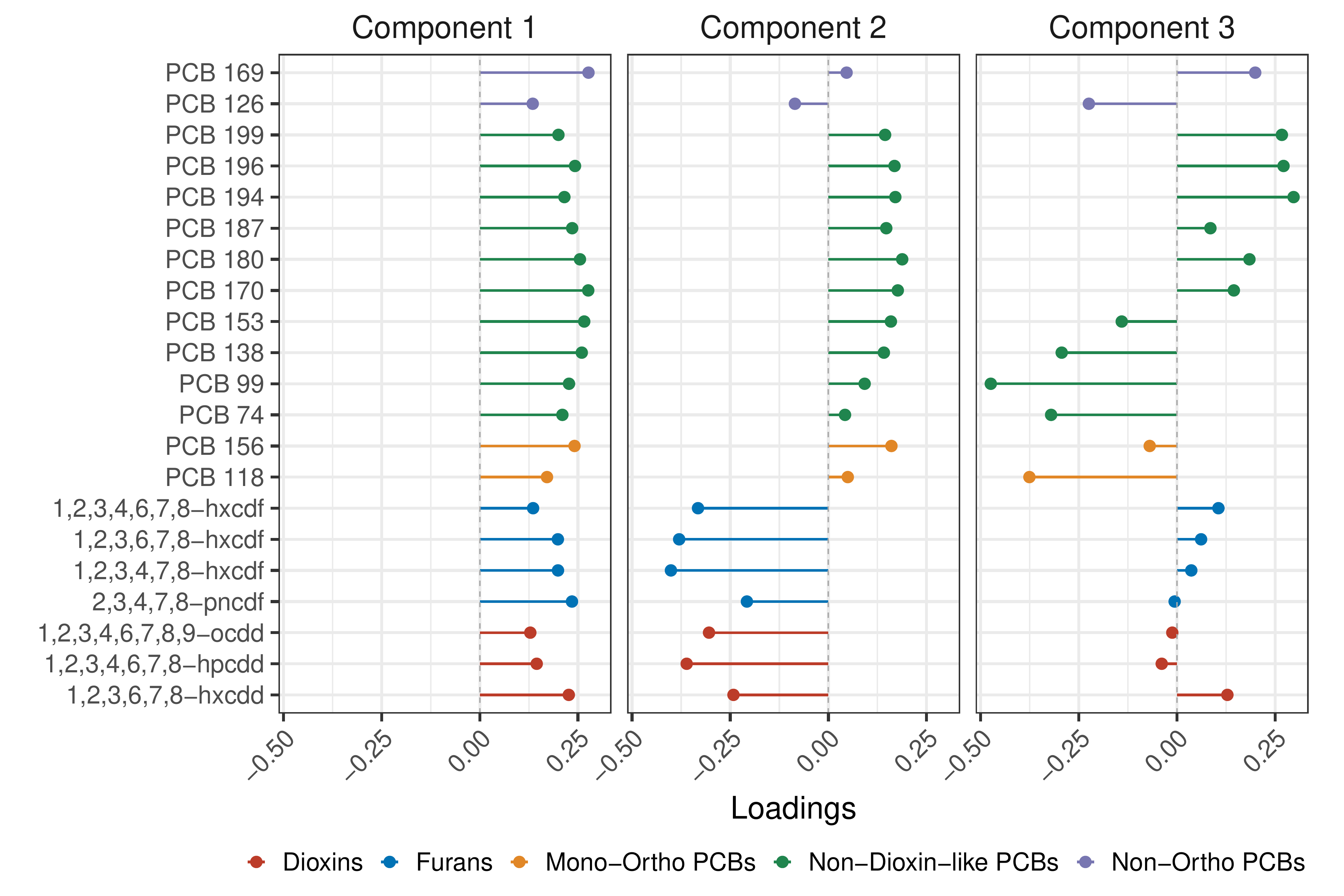}
   \caption{SVD-identified components of PCP-LOD $\hat{L}$ matrix of underlying POP exposure in 2001--2002 NHANES participants. PCP-LOD chose a three rank solution based on random hold-out cross-validation.}
    \label{fig:nhanes_load}
\end{figure}

We applied PCP-LOD to identify underlying patterns of POP exposure and extreme exposure events that were not explained by these patterns without making \textit{a priori} assumptions concerning the number of patterns or sparse events. PCP-LOD returned a low-rank matrix of rank three, which corresponds with three patterns of POP exposure in the $\hat{L}$ matrix. Figure~\ref{fig:nhanes_corr_l} depicts $\hat{L}$'s correlation matrix along side the correlation matrix of the raw data (Figure~\ref{fig:nhanes_corr}). By removing sparse events and residual noise, PCP-LOD increased the correlations between POPs. To characterize underlying patterns, we extracted principal components from the low-rank matrix using SVD.

The three components distinguished by PCP-LOD included one component of overall POP exposure, a component that separated dioxins and furans from PCBs, and a third component that separated higher molecular weight PCBs from lower molecular weight PCBs (Figure~\ref{fig:nhanes_load}). The first component explained 79.4\% of the variance in the low-rank matrix, the second explained 14.6\%, and the third explained 6.0\%.

\begin{figure}
    \centering
\includegraphics[width=0.95\textwidth]{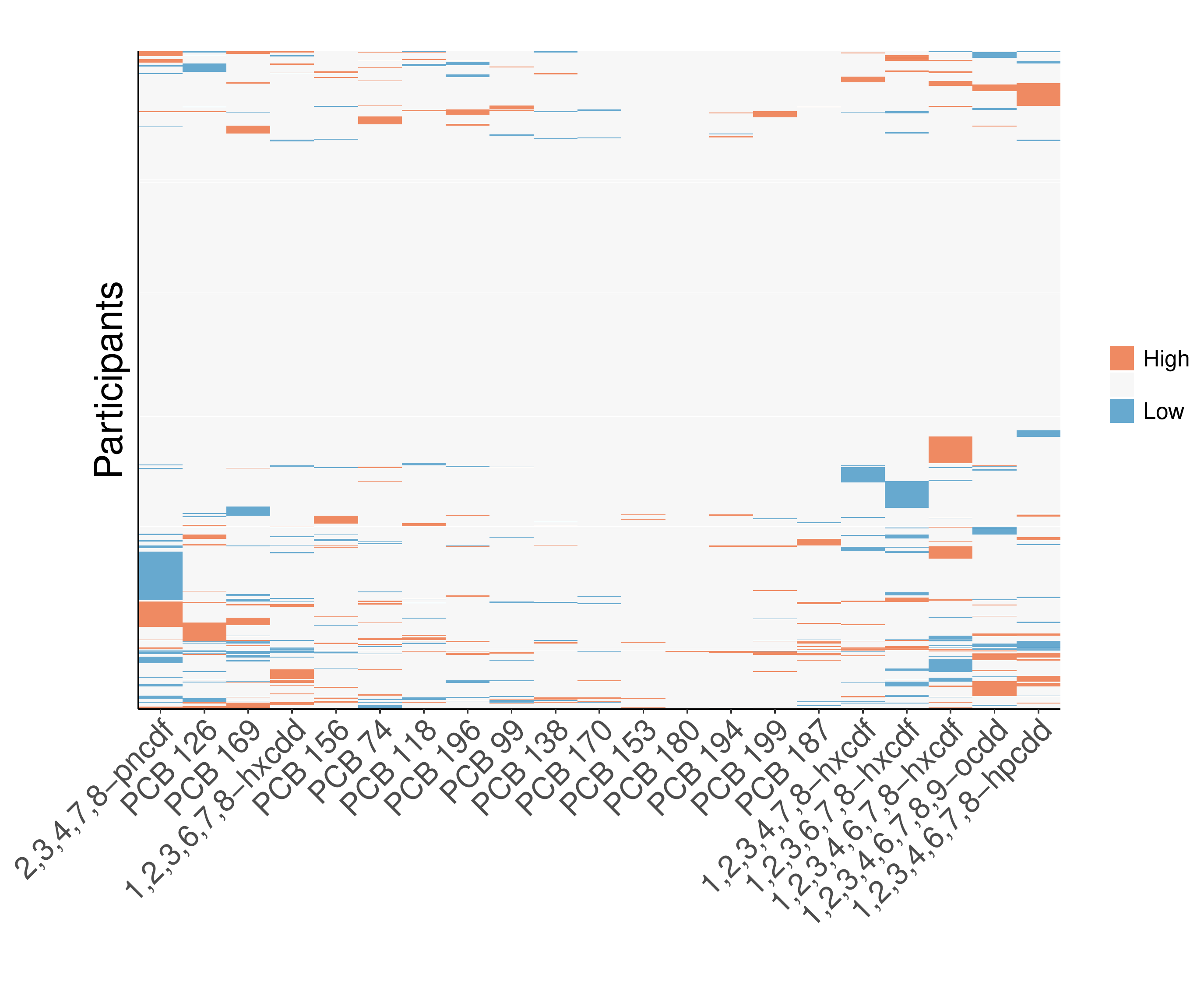}
   \caption{PCP-LOD solution $\hat{S}$ matrix of sparse events of POP exposure in 2001--2002 NHANES participants. To facilitate visualization, we have categorized sparse values into high and low exposure events. Red indicates an extremely high exposure event $\left(> 2 \times \sqrt{\mathrm{Var}(X_{p}^{\mathrm{obs}} - L_{p}^{\mathrm{obs}})}\right)$; blue indicates an extremely low exposure event $\left(< -2 \times \sqrt{\mathrm{Var}(X_{p}^{\mathrm{obs}} - L_{p}^{\mathrm{obs}})}\right)$. White indicates sparsity. POPs (columns) and NHANES participants (rows) are hierarchically clustered to further facilitate visualization.}
    \label{fig:sparse}
\end{figure}

PCP-LOD partitioned the variation that was unexplained by the low-rank structure into a sparse matrix of large outlying values and the remaining residuals. The $\hat{S}$ matrix contained mostly zero values, with 5.7\% of entries being non-sparse. Sparse observations were generally weakly correlated, with the absolute value of r $<$ 0.15 for 70\% of Spearman correlations between sparse chemical exposure events. Table~\ref{tab:sparse} summarizes the number of individuals with uniquely high $\left(> 2 \times \sqrt{\mathrm{Var}(X_{p}^{\mathrm{obs}} - L_{p}^{\mathrm{obs}})}\right)$ or low $\left(< -2 \times \sqrt{\mathrm{Var}(X_{p}^{\mathrm{obs}} - L_{p}^{\mathrm{obs}})}\right)$ exposure events. Figure~\ref{fig:sparse} describes participant-specific sparse events. Most participants had no extreme exposures (44\%) or only extremely low exposures (18\%). Twenty-two percent had one high unique event on a single chemical, and 16\% had between two and six high exposures across 21 chemicals left unexplained by the identified patterns.
  
\begingroup
\renewcommand{\arraystretch}{1.1}
\begin{table} \centering 
  \caption{Summary of extreme events captured in the sparse component of PCP-LOD from a mixture of 21 POPs measured in 1,000 participants in NHANES 2001--2002. Entries are counts of participants with uniquely low and/or high events, organized by row and column, respectively.}
  \label{tab:sparse} 
\begin{tabular}{@{\extracolsep{5pt}} c|r|rrrrrrr|r} 
\hline 
\hline 
& & \multicolumn{7}{c}{High Unique Events} \\
\hline
 & & 0 & 1 & 2 & 3 & 4 & 5 & 6 & $+^*$ \\ 
\hline
\multirow{8}{*}{\rotatebox[origin=c]{90}{Low Unique Events}}& 0 & 439 & 141 & 41 & 11 & 5 & 0 & 1 & 638\\ 
& 1 & 147 & 46 & 30 & 13 & 5 & 1 & 0 & 242 \\ 
& 2 & 27 & 19 & 11 & 6 & 3 & 1 & 0 & 67 \\ 
& 3 & 8 & 11 & 9 & 8 & 2 & 1 & 1 & 40 \\ 
& 4 & 0 & 1 & 2 & 3 & 2 & 0 & 1 & 9 \\ 
& 5 & 0 & 0 & 0 & 1 & 0 & 0 & 1 & 2 \\ 
& 6 & 0 & 0 & 0 & 0 & 1 & 1 & 0 & 2 \\ 
\cline{2-10}
& $+^\dagger$ & 621 & 218 & 93 & 42 & 18 &  4 & 4 & 1000 \\ 
\hline
\hline
\multicolumn{10}{l}{$+^\dagger$ Column sums of uniquely high events.} \\
\multicolumn{10}{l}{$+^*$ Row sums of uniquely low events.} \\
\end{tabular} 
\end{table} 
\endgroup

PCA conducted on the POP mixture chose three components that explained $\ge$ 80\% of the variance and returned loadings and scores much the same as those from $\hat{L}$ (results not shown). Using the three chosen components, PCA's relative prediction error on values $>$ LOD was 0.30, similar to PCP-LOD's relative error of 0.32 when comparing only $\hat{L}$ with the original data. However, when including $\hat{S}$ in the solution ($\hat{L} + \hat{S}$), PCP-LOD's relative error on values $>$ LOD was 0.07. This is more comparable to the PCA solution when including all 21 components, 0.06, which does not accomplish any dimension reduction.

\section{Discussion}
We propose PCP-LOD as a new approach to identify patterns---and extreme events left unexplained by patterns---underlying environmental chemical mixtures in the presence of values $<$ LOD. Our simulation studies highlighted three main advantages of PCP-LOD over PCA at identifying patterns in environmental mixtures: (1) reduced error in estimated patterns of exposure, (2) identification of extreme or unique events, and (3) improved estimation of values $<$ LOD.

Patterns identified by PCP-LOD are additionally more robust to noise and incomplete data than more traditional pattern identification methods because patterns in $L$ are not influenced by events in $S$. PCP-LOD estimated the underlying low-rank structure of $L$ with lower relative error than PCA under all realistic simulation scenarios. PCA outperformed PCP-LOD for two error structures when 75\% of the dataset was simulated as $<$ LOD. In this case, PCP-LOD used 25\% to re-construct 75\% of the data, and poorer performance was expected. However, it is unlikely that an environmental health researcher will face a chemical mixture with 75\% of all values $<$ LOD. In our application to POPs detected in over 50\% of measurements among NHANES participants, 76\% of all observations were $>$ LOD. In the entire POP mixture of 34 chemicals, with five chemicals never detected, 52\% of all observations were $>$ LOD. We observed the highest relative prediction error across all simulations for values $<$ LOD in simulated datasets. This held for PCA, as well, and applies to all methods to address censored or missing data.

In our simulations and application to NHANES data, we did not make use of the non-negativity constraint, as SVD returns solutions with negative values. However, we paired PCP-LOD with SVD to make results comparable with those of PCA. This is not a constraint of PCP-LOD, as it may be paired with various dimension reduction techniques. Because of the non-negativity constraint on the $L$ matrix, for example, PCP-LOD can be paired with NMF to provide results interpretable on an additive scale with a parts-based representation. \citep{lee1999learning}.

The three components underlying the NHANES mixture distinguished by PCP-LOD represent one pattern of exposure to all POPs and two patterns grouped by known structural and toxicological properties. More than 90\% of human exposure to PCBs, dioxins, and furans is through the food supply, mainly meat, dairy, and seafood \cite{faroon2000toxicological, world1999dioxins, loganathan2020pcbs}. Thus, the first component of comprehensive exposure may be interpreted as a dietary source of these POPs. The second component separated dioxins and furans, which are generally more toxic, from PCBs \cite{iarc2012review}. Accordingly, we can understand the second component as a measure of toxicity. The third component separated lower molecular weight PCBs from higher molecular weight PCBs, where larger numbers indicate more chlorine atoms and larger molecules. Higher chlorinated congeners tend to bioaccumulate more than lower chlorinated congeners \cite{steele1986estimates, hopf2009background}. Depending on the research question, any or all of these components could be included in subsequent analyses with health outcomes.

In the original POP mixture, individuals with high values on any chemicals were likely to have high values on other chemicals, or equivalently, individuals with low values on any chemicals were likely to have low values on other chemicals. PCP-LOD captured this in a component representing overall mixture exposure. After removing the underlying patterns in the mixture described in $\hat{L}$, high (or low) exposure events on individual chemicals did not indicate high (or low) exposure to other chemicals, i.e., sparse events in $\hat{S}$ were not highly correlated. About half (51\%) of the unique low exposure events were $<$ LOD in the original mixture; these values $<$ LOD were not explained by overall low exposure or by the other identified patterns.

The ability to identify and separate extreme events is a unique feature offered by PCP-LOD and cannot be found in other methods. These unique or extreme events not captured in $L$ may themselves be risk factors (e.g., wildfires---unique events not explained by commonly recognized air pollution sources---for asthma emergency admissions)\cite{delfino08}, or they may modify an association with one of the $L$ components (e.g., a Saharan dust episode might modify the association with traffic-related pollution) \cite{karanasiou12}. Next steps could entail including $S$ exposures along with identified patterns from $L$ in a health model with some form of penalization (e.g., lasso or elastic net).

While PCP-LOD addresses several drawbacks of existing methods, it does not overcome all limitations of pattern identification in environmental mixtures. First, in multi-pollutant exposures the `true' originating mechanism is almost never known, thus PCP-LOD cannot provide the `correct' answer. PCP-LOD, like other methods employed in our field, should be used in conjunction with subject area expertise. The interpretability of results relies on this expert knowledge. This limitation applies, however, to all methods to address research questions concerning patterns of environmental exposures. Second, including scores obtained from any dimension reduction technique paired with PCP-LOD in a health model ignores the uncertainty inherent in the solution selection, resulting in underestimated confidence intervals and, potentially, spurious results \cite{mak14_unc}. Third, some datasets will likely be high-dimensional, with a large number of correlated chemical measurements for each participant. In this situation, PCP-LOD still performs well, provided the rank of the target matrix $L_{0}$ is small enough compared to $n$ (e.g., $r<c p / \log ^{2} n$, where $c$ is a constant) \cite{zhou2010stable}. Additionally, our application findings should be interpreted in light of their limitations. First, as is the case when using chemical biomarkers, our study is susceptible to exposure measurement error. In a noisy setting, any method will exhibit an inaccuracy in the estimated left singular vectors, which is commensurate with the noise level. Nevertheless, even in this setting, the results produced by PCP-LOD are stable with respect to noise \cite{zhou2010stable}. Second, our results may not be generalizable beyond the study population. While NHANES includes a nationally-representative sample of the general non-institutionalized US population \cite{johnson2013national}, we did not account for the complex sampling design and weights of the study \cite{curtin2012national}. Thus, the PCP-LOD-identified patterns may represent sources or behaviors distinct to the participants. 

PCP-LOD also has numerous strengths when compared with existing methods to identify exposure patterns in environmental mixtures, which require strong assumptions and have key limitations. As a consequence, their use has resulted in heterogeneous and inconsistent findings across studies \cite{gibson2019complex}. Moreover, results from methods that are not generalizable or interpretable hinder their use in the design and development of regulations, policies, and targeted interventions. Original PCP has few assumptions, namely that $L$ is not sparse and that $S$ is not low-rank \cite{candes2011robust}. This is an appealing feature of a tool when the underlying truth is not known. PCP-LOD directly addresses several additional limitations of existing methods: (1) its solution is not necessarily orthogonal, allowing correlations between patterns, (2) its solution is non-negative, so patterns can exist in an interpretable space, (3) its parameters do not require tuning by the researcher, meaning that the choice of number of patterns in $L$ is not subjective, and (4) PCP-LOD is robust to extreme values because of the novel $S$ matrix.

To our knowledge, this work represents the first instance of decomposing the structure among chemicals in an additive manner. By separating the unique events from underlying patterns, PCP-LOD provides the opportunity to include extreme events in analyses, where they previously may have been suppressed or discarded. The theory-backed parameter selection and cross-validation enhances reproducibility of PCP-LOD, ensuring that two different research groups with the same dataset will identify the same optimal number of patterns. PCP-LOD may be employed when environmental epidemiologists have research questions concerning sources or behaviors leading to chemical exposure or patterns underlying exposure to multi-pollutant mixtures, especially when data are noisy, incomplete, or may contain extreme exposure events.

\section*{Acknowledgements}

This work was partially supported by the National Institutes of Environmental Health (NIEHS) individual fellowship grant F31 ES030263, as well as PRIME R01 ES028805 and P30 ES009089.

\clearpage
\setcounter{figure}{0}
\setcounter{table}{0}
\renewcommand{\thefigure}{S\arabic{figure}}
\renewcommand{\thetable}{S\arabic{table}}
\section{Supplemental Materials}

\begingroup
\renewcommand{\arraystretch}{1.2}
\begin{table}[!h]
\centering
\caption{Overall relative prediction error comparing PCA and PCP-LOD solutions with simulated data before noise or sparse events were added. Values represent $25^{th}$, $50^{th}$, and $75^{th}$ percentiles of error distribution over 200 simulations (100 for $p$ = 16; 100 for $p$ = 48).}
\label{table:supp1}
\begin{tabular}{llrrr|rrr|rrr}
  \hline
  \hline
 & & \multicolumn{3}{c}{25\% $<$ LOD} & \multicolumn{3}{c}{50\% $<$ LOD} & \multicolumn{3}{c}{75\% $<$ LOD} \\ 
Simulated Data & Method & $25^{th}$ & $50^{th}$ & $75^{th}$ & $25^{th}$ & $50^{th}$ & $75^{th}$ & $25^{th}$ & $50^{th}$ & $75^{th}$ \\
\hline
Low noise  & PCA & 0.27 & 0.30 & 0.31 & 0.28 & 0.31 & 0.32 & 0.36 & 0.38 & 0.39 \\  
 & PCP-LOD & 0.07 & 0.08 & 0.10 & 0.12 & 0.13 & 0.14 & 0.24 & 0.26 & 0.27 \\ 
\hline  
High noise & PCA & 0.40 & 0.43 & 0.45 & 0.40 & 0.42 & 0.45 & 0.55 & 0.57 & 0.59 \\  
 & PCP-LOD & 0.24 & 0.30 & 0.37 & 0.35 & 0.38 & 0.42 & 0.63 & 0.67 & 0.71 \\  
\hline  
Low noise & PCA & 0.15 & 0.21 & 0.31 & 0.19 & 0.23 & 0.33 & 0.35 & 0.37 & 0.44 \\ 
+ sparse events & PCP-LOD & 0.08 & 0.09 & 0.12 & 0.15 & 0.16 & 0.17 & 0.36 & 0.38 & 0.40 \\ 
\hline  
\hline
\end{tabular}
\end{table}
\endgroup

\begingroup
\renewcommand{\arraystretch}{1.2}
\begin{table}[!h]
\centering
\caption{Relative prediction error on observations $>$ LOD comparing PCA and PCP-LOD solutions with simulated data before noise or sparse events were added. Values represent $25^{th}$, $50^{th}$, and $75^{th}$ percentiles of error distribution over 200 simulations (100 for $p$ = 16; 100 for $p$ = 48).}
\label{table:supp2}
\begin{tabular}{llrrr|rrr|rrr}
  \hline
  \hline
 & & \multicolumn{3}{c}{25\% $<$ LOD} & \multicolumn{3}{c}{50\% $<$ LOD} & \multicolumn{3}{c}{75\% $<$ LOD} \\ 
Simulated Data & Method & $25^{th}$ & $50^{th}$ & $75^{th}$ & $25^{th}$ & $50^{th}$ & $75^{th}$ & $25^{th}$ & $50^{th}$ & $75^{th}$ \\
\hline
Low noise  & PCA & 0.25 & 0.27 & 0.29 & 0.23 & 0.25 & 0.26 & 0.21 & 0.23 & 0.25 \\  
 & PCP-LOD & 0.06 & 0.07 & 0.09 & 0.06 & 0.07 & 0.08 & 0.05 & 0.05 & 0.07 \\ 
\hline  
High noise & PCA & 0.39 & 0.41 & 0.43 & 0.35 & 0.38 & 0.41 & 0.31 & 0.35 & 0.36 \\  
 & PCP-LOD & 0.23 & 0.29 & 0.36 & 0.27 & 0.32 & 0.38 & 0.33 & 0.36 & 0.41 \\  
\hline  
Low noise & PCA & 0.13 & 0.19 & 0.29 & 0.12 & 0.18 & 0.27 & 0.12 & 0.18 & 0.27 \\ 
+ sparse events & PCP-LOD & 0.06 & 0.08 & 0.11 & 0.07 & 0.08 & 0.11 & 0.11 & 0.12 & 0.14 \\ 
\hline  
\hline
\end{tabular}
\end{table}
\endgroup

\begingroup
\renewcommand{\arraystretch}{1.2}
\begin{table}[!h]
\centering
\caption{Relative prediction error on observations $<$ LOD comparing PCA and PCP-LOD solutions with simulated data before noise or sparse events were added. Values represent $25^{th}$, $50^{th}$, and $75^{th}$ percentiles of error distribution over 200 simulations (100 for $p$ = 16; 100 for $p$ = 48).}
\label{table:supp3}
\begin{tabular}{llrrr|rrr|rrr}
  \hline
  \hline
 & & \multicolumn{3}{c}{25\% $<$ LOD} & \multicolumn{3}{c}{50\% $<$ LOD} & \multicolumn{3}{c}{75\% $<$ LOD} \\ 
Simulated Data & Method & $25^{th}$ & $50^{th}$ & $75^{th}$ & $25^{th}$ & $50^{th}$ & $75^{th}$ & $25^{th}$ & $50^{th}$ & $75^{th}$ \\
\hline
Low noise  & PCA & 0.92 & 1.04 & 1.17 & 0.78 & 0.85 & 0.93 & 0.82 & 0.85 & 0.90 \\  
 & PCP-LOD & 0.36 & 0.37 & 0.38 & 0.51 & 0.52 & 0.54 & 0.68 & 0.70 & 0.73 \\ 
\hline  
High noise & PCA & 0.67 & 0.72 & 0.75 & 0.64 & 0.67 & 0.70 & 0.99 & 1.01 & 1.03 \\  
 & PCP-LOD & 0.39 & 0.45 & 0.52 & 0.66 & 0.70 & 0.73 & 1.10 & 1.25 & 1.36 \\  
\hline  
Low noise & PCA & 0.65 & 0.77 & 1.03 & 0.67 & 0.74 & 0.88 & 0.84 & 0.89 & 0.95 \\ 
+ sparse events & PCP-LOD & 0.41 & 0.42 & 0.44 & 0.60 & 0.62 & 0.63 & 0.88 & 0.97 & 1.03 \\ 
\hline  
\hline
\end{tabular}
\end{table}
\endgroup
\clearpage

\begin{figure}
    \centering
\includegraphics[width=.75\textwidth]{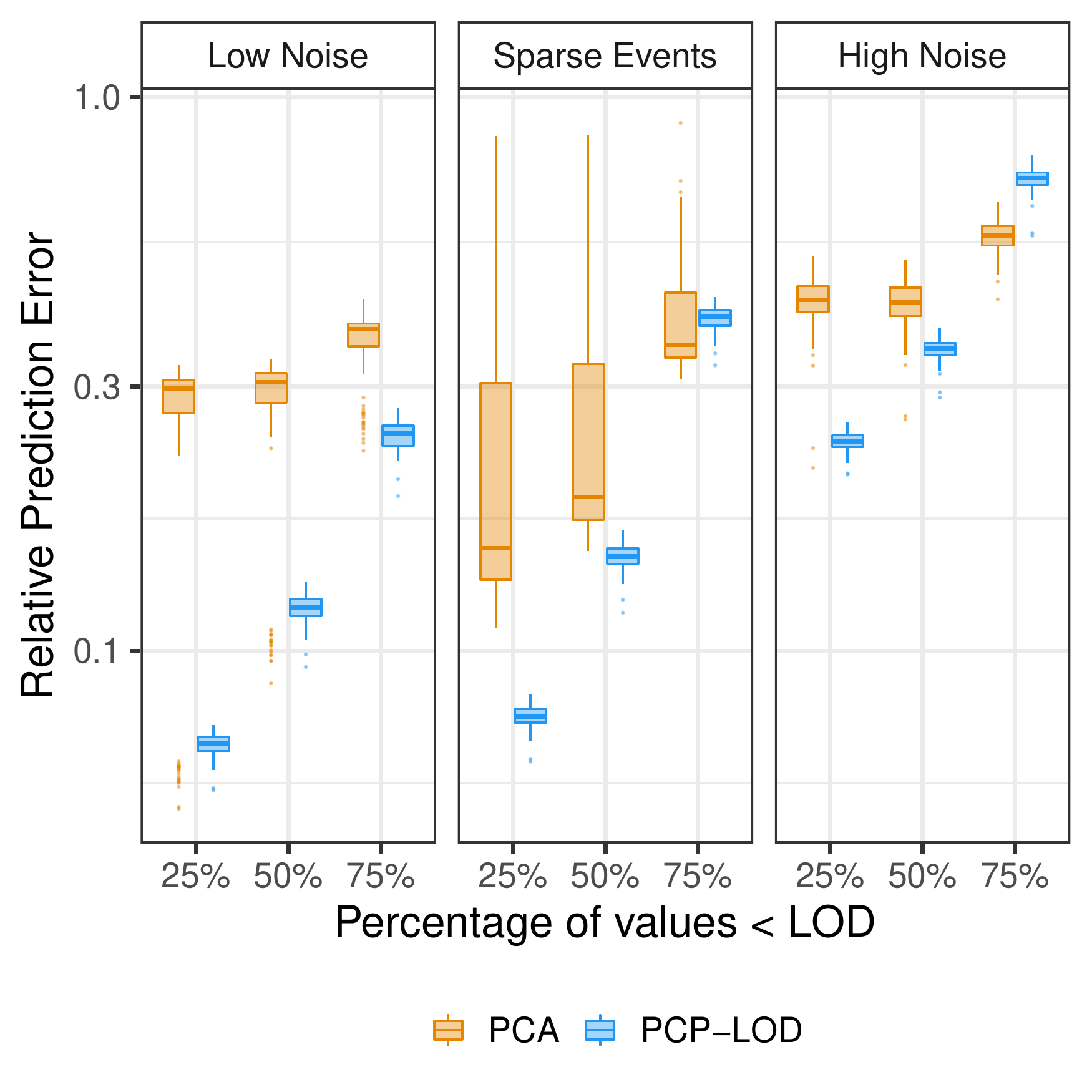}
   \caption{Overall relative predictive error of PCP-LOD and PCA on simulated data with $p$ = 48 across increasing proportions of data below the limit of detection. The panels show results for different structures of added noise. Box plots display summary statistics for each method across 100 simulations. The bottom and top hinges of the boxes correspond to the first and third quartiles (the 25$^{th}$ and 75$^{th}$ percentiles), respectively. The upper (lower) whiskers extend from the hinge to the largest (smallest) value no further than 1.5 $\times$ IQR from the hinge (where IQR is the inter-quartile range, or distance between the first and third quartiles).}
    \label{fig:overall_48}
\end{figure}

\begin{figure}
    \centering
\includegraphics[width=.75\textwidth]{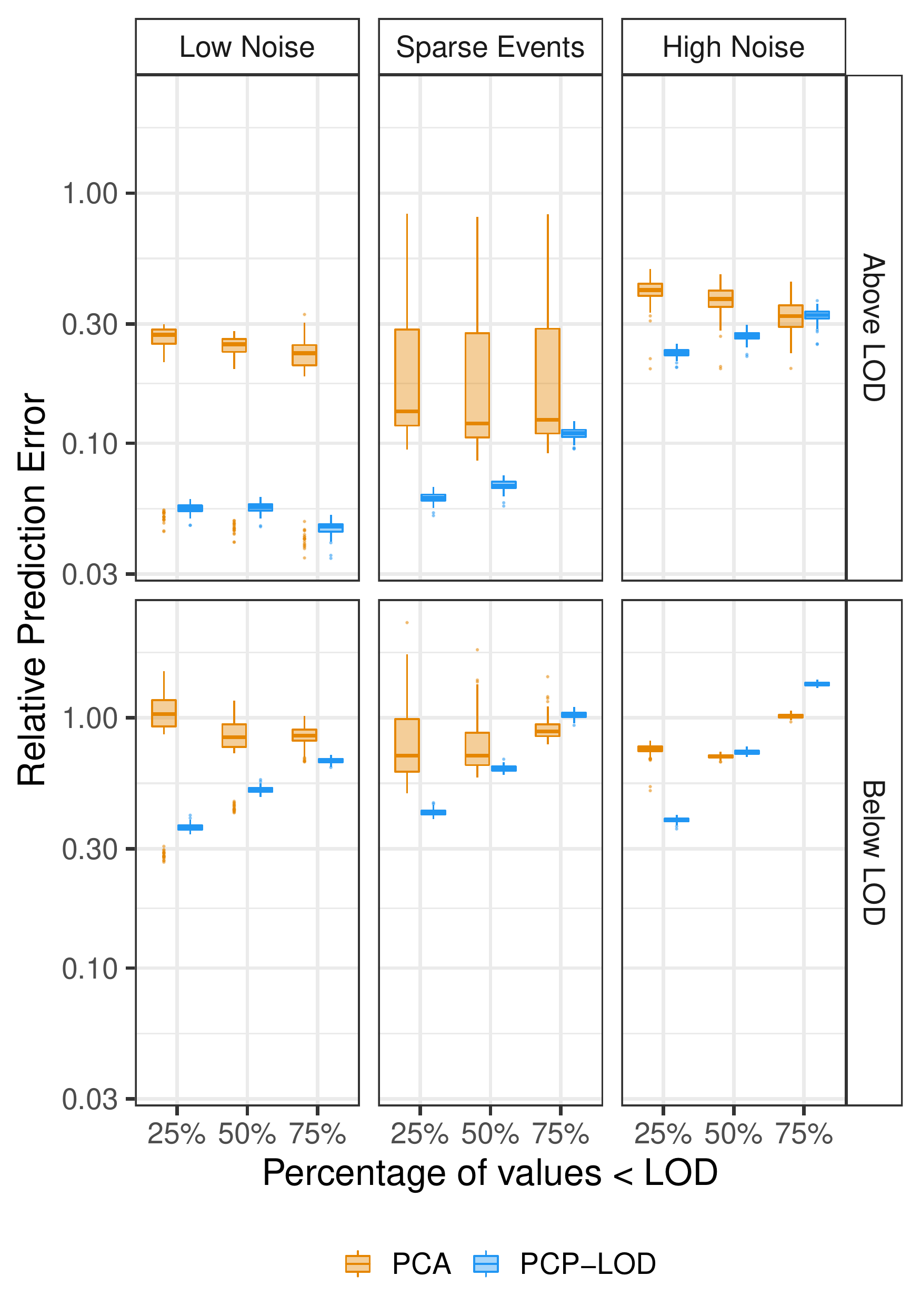}
   \caption{Relative predictive error of PCP-LOD and PCA on simulated data with $p$ = 48 stratified by detection. The panel columns separate results from different structures of added noise, and the panel rows separate values that were simulated as observed (top row) from those simulated as below the limit of detection (bottom row). Box plots display summary statistics for each method across 100 simulations.}
    \label{fig:above_below_48}
\end{figure}

\begin{figure}
    \centering
\includegraphics[width=.75\textwidth]{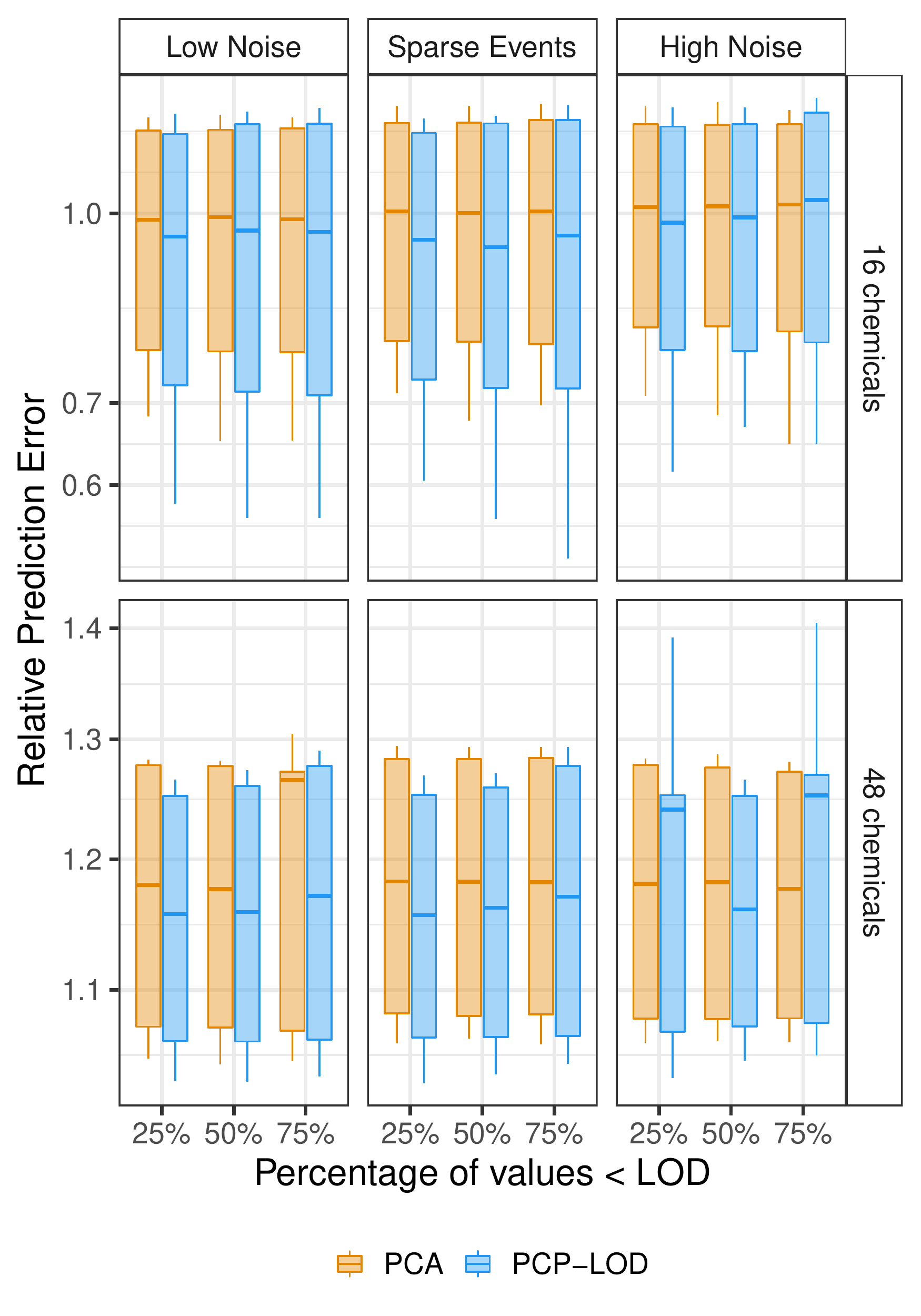}
   \caption{Relative estimation error of PCP-LOD and PCA solution chemical loadings (i.e., right eigenvectors) compared with those of the simulated data before noise was added. The panel columns separate results from different structures of added noise, and panel rows present two simulated mixture sizes. Box plots display summary statistics for each method across 100 simulations.}
    \label{fig:svd_right}
\end{figure}

\clearpage
\bibliographystyle{unsrtnat}
\bibliography{pcplod}

\end{document}